\documentclass[prd,aps,showpacs,tightenlines]{revtex4}  %%preprint,
\usepackage{mathrsfs}
\usepackage{amsmath}
\usepackage{amssymb}
\usepackage{epsfig}
\usepackage{graphicx}
\usepackage{booktabs}
\usepackage{multirow}

\begin{document}
%%%%%%%%%%%%%%%%%%%%%%%%%%%%%%%%%%%%%%%%%%%%%
%\renewcommand{\arraystretch}{0.5}
\newcommand{\psl}{ p \hspace{-1.8truemm}/ }
\newcommand{\nsl}{ n \hspace{-2.2truemm}/ }
\newcommand{\vsl}{ v \hspace{-2.2truemm}/ }
\newcommand{\epsl}{\epsilon \hspace{-1.8truemm}/\,  }

%%%%%%%%%%%%%%%%%%%%%%%%%%%%%%%%%%%%%%%%%%%%%%%%%%%%%%

\title{The    $B_c\rightarrow \psi(2S)\pi$, $\eta_c(2S)\pi$ decays in the perturbative QCD approach}
\author{Zhou Rui$^1$}\email{zhourui@ncst.edu.cn}
\author{Wen-Fei Wang$^{2,3}$}
\author{Guang-xin Wang$^1$}
\author{Li-hua Song$^1$}
\author{Cai-Dian L\"u$^3$}
\affiliation{$^1$College of Sciences, North China University of Science and Technology,
                 Tangshan 063009,  China, \\
            $^2$Department of Physics and Institute of Theoretical Physics,
                Shanxi University, Taiyuan, Shanxi 030006, China,\\
            $^3$Center for Future High Energy Physics, Institute of High Energy Physics,
                Chinese Academy of Sciences, Beijing 100049, China}
\date{\today}

\begin{abstract}
Nonleptonic two body $B_c$ decays including radially excited  $\psi(2S)$ or $\eta_c(2S)$ mesons
in the final state are studied   using the perturbative QCD approach based on $k_T$ factorization. The charmonium distribution
amplitudes are extracted from the $n = 2, l = 0$ Schr$\ddot{o}$dinger states for the harmonic
oscillator potential. Utilizing these distribution amplitudes, we  calculate the numerical results of the $B_c\rightarrow \psi(2S),\eta_c(2S)$
transition form factors and branching fractions of $B_c\rightarrow \psi(2S)\pi, \eta_c(2S)\pi$
decays. The ratio between two decay modes $B_c\rightarrow \psi(2S)\pi$ and $B_c\rightarrow J/\psi\pi$
is compatible with the experimental data within uncertainties, which indicates that the  harmonic-oscillator wave functions for $\psi(2S)$ and $\eta_c(2S)$ work well. It is found that the branching fraction of $B_c\rightarrow \eta_c(2S)\pi$,
which is dominated by the twist-3 charmonium distribution amplitude, can reach the order of $10^{-3}$.
We hope it can be measured soon     in the   LHCb experiment.
\end{abstract}



\maketitle

\section{Introduction}

The meson $B_c$, a pseudoscalar ground state of $b$ and $c$ quarks, can only decay through weak
interactions. Either of the heavy quarks ($b$ or $c$) in it can decay individually, which makes it
an ideal system to study weak decays of heavy quarks. Around $\mathcal{O}(10^{9})$ mesons can be
anticipated with $1$ fb$^{-1}$ of data at the LHC~\cite{cpl061302}, which is sufficient for studying
the $B_c$ meson family systematically. Up to now, several new decay channels of the $B_c$ meson
\cite{prl281802,jhep075,prd112012,jhep094,prl181801} have been successfully observed by the LHCb
Collaboration, while an excited $B_c$  meson state which is consistent with expectations of the
$B_c(2S)$ has been found by the ATLAS detector~\cite{14071032}.

Recently, the LHCb Collaboration observed the decay mode $B_c\rightarrow \psi(2S)\pi$ for the
first time with the measured ratio of the branching fractions as \cite{prd071103}
\begin{eqnarray}
\frac{\mathcal {B}(B_c\rightarrow \psi(2S)\pi)}{\mathcal {B}(B_c\rightarrow J/\psi\pi)}
=0.250 \pm 0.068 (\text{stat}) \pm0.014 (\text{syst})\pm 0.006(\mathcal {B})\;.
\end{eqnarray}
The last term above accounts for the uncertainty on
$\mathcal {B}(\psi(2S)\rightarrow \mu^+\mu^-)/\mathcal {B}(J/\psi\rightarrow \mu^+\mu^-)$.
Although there is not much data for the $B_c$ meson decaying into two-body final states containing
a radially excited charmonium such as $ \psi(2S)$ or $\eta_c(2S)$ except the $B_c\rightarrow \psi(2S)\pi$
channel, many theoretical studies of nonleptonic $B_c$ decays with radially excited
charmonium mesons in the  final state have been performed by using various approaches.
For example, in Ref.~\cite{prd017501},
the authors computed the branching ratios for $B_c\rightarrow \psi(2S)X$
decays with the modified harmonic-oscillator wave function in the light front quark model;
in Ref.~\cite{11022190}, the ISGW2 quark model was adopted to study the production of radially excited
charmonium mesons in two-body nonleptonic $B_c$ decays, the relativistic (constituent) quark
model, the potential model, the QCD relativistic potential model, and the improved instantaneous BS
equation and Mandelstam approach were adopted in Refs.~\cite{prd094020,prd4133}, Ref.~\cite{prd3399},
Ref~\cite{prd034012} and Ref~\cite{14113428}, respectively.
However, all of these calculations are based on a naive factorization hypothesis, with various form
factor inputs. There are uncontrolled large theoretical errors with quite different numerical results,
and  most of them cannot give any theoretical error estimates because of the unreliability
of these models.

The perturbative QCD approach (pQCD) \cite{prl4388} based on $k_T$ factorization, which not only can deal with the emission
diagrams corresponding to the naive factorization terms basically, but  can also handle
well the nonfactorizable diagrams by introducing the wave function of the light meson in the final
states of the $B_c$ decay modes, is widely used in the nonleptonic two-body $B_c$
decays~\cite{prd014022,prd074017,prd074012,prd054029,jpg035009,prd074033,14010151,epjc45711,
epjc63435,prd074008,prd074019,prd074027}. In our recent work~\cite{14075550}, the pQCD approach was
used successfully in describing the S-wave ground state charmonium decays of $B_c$ meson
based on the harmonic-oscillator wave functions for the charmonium 1S states.
In this work, we will use the harmonic-oscillator wave function as the approximate wave function
of the 2S states and study the $B_c\rightarrow \psi(2S)\pi, \eta_c(2S)\pi$ decays in the pQCD
approach to provide a ready reference to the existing and forthcoming experiments.

The structure of this paper is organized as follows. After this Introduction, we describe the
wave functions of radially excited charmonium mesons $\psi(2S),\eta_c(2S)$ in Sect.~II.
We calculate and present the expressions for the $B_c\rightarrow\psi(2S),\eta_c(2S)$ transition
form factors in the large-recoil regions and the $B_c\rightarrow\psi(2S)\pi,\eta_c(2S)\pi$ decay
amplitudes in Sect.~III. The numerical results and relevant discussions are given in Sect.~IV, and
Sect.~V contains a brief summary.

\section{wave functions}\label{sec:f-work}

In hadronic B decays, there are several energy scales involved. In the expansion of the inverse power of heavy quark mass, the hadronic matrix element can be factorized into perturbative and nonperturbative factors. In the pQCD approach, the decay amplitude ${\cal A}(B_c \to M_2 M_3)$ can be written conceptually
as the convolution \cite{prl4388}
\begin{eqnarray}
{\cal A}(B_c \to M_2 M_3)\sim \int\!\! d^4k_1 d^4k_2 d^4k_3\ \mathrm{Tr}
\left [ C(t) \Phi_{B_c}(k_1) \Phi_{M_2}(k_2) \Phi_{M_3}(k_3)H(k_1,k_2,k_3, t) \right ],
\label{eq:factorization}
\end{eqnarray}
where $k_i$'s are momenta of spectator quarks included in each meson, and $``\mathrm{Tr}"$ denotes
the trace over Dirac and color indices. In the above convolution, $C(t)$ is the Wilson coefficient
evaluated at scale $t$, the function $H(k_1,k_2,k_3,t)$ describes the four quark operator and the
spectator quark connected by a hard gluon, which can be perturbatively calculated including all
possible Feynman diagrams without end-point singularity. The wave functions $\Phi_{B_c}(k_1)$, $\Phi_{M_2}$
and $\Phi_{M_3}$ describe the hadronization of the quark and anti-quark in the $B_c$ meson, the
charmonium meson $\psi(2S)$ or $\eta_c(2S)$ and the final state light meson pion, respectively.

As a heavy quarkonium    discussed in Refs.~\cite{14075550,prd90-094018}, the nonrelativistic QCD framework can be
applied for the $B_c$ meson, which means its leading-order wave function should be just the zero-point
wave function with the distribution amplitude
\begin{eqnarray}
\label{eq-Bc-wave}
\phi_{B_c}(x)= \frac{f_{B_c}}{2\sqrt{2N_c}}\delta(x-m_c/m_{B_c})
\exp[-\omega^2_{B_c}b^2/2]\;.
\end{eqnarray}
For the light meson pion, we adopt the same distribution amplitudes $\phi_{\pi}^A(x)$ and
$\phi_{\pi}^{P,T}(x)$ as defined in Refs.~\cite{ball1999,ball2006}.

The harmonic-oscillator wave functions has been adopted to describe the 1S state mesons
\cite{epjc60107,prd037501,prd114019}, and they can explain the experimental data well~\cite{14075550}.
In the quark model,  $\eta_c(2S)$ and $\psi(2S)$ are the first excited states of $\eta_c$
and $J/\psi$, respectively.  The 2S means that for these states, the principal quantum number $n=2$ and
the orbital angular momentum $l=0$. The definitions of the 2S state wave functions are similar
to the 1S states via the nonlocal matrix elements \cite{prd114008},
\begin{eqnarray}\label{eq:wavve}
\langle \psi(2S) (P, \epsilon^L)|\bar{c}(z)_{\alpha}c(0)_{\beta}|0\rangle
&=& \frac{1}{\sqrt{2N_c}}\int_0^1 dxe^{ixP\cdot z}
[m\rlap{/}{\epsilon^L}_{\alpha\beta}\psi^L(x,b)+(\rlap{/}{\epsilon^L}\rlap{/}{P})_{\alpha\beta}\psi^t(x,b)], \nonumber\\
\langle \psi(2S) (P, \epsilon^T)|\bar{c}(z)_{\alpha}c(0)_{\beta}|0\rangle
&=& \frac{1}{\sqrt{2N_c}}\int_0^1 dxe^{ixP\cdot z}
[m\rlap{/}{\epsilon^T}_{\alpha\beta}\psi^V(x,b)+(\rlap{/}{\epsilon^T}\rlap{/}{P})_{\alpha\beta}\psi^T(x,b)],\nonumber\\
\langle \eta_c(2S)(P)|\bar{c}(z)_{\alpha}c(0)_{\beta}|0\rangle &=& -\frac{i}{\sqrt{2N_c}}\int_0^1 dxe^{ixP\cdot z}
[(\gamma_5\rlap{/}{P})_{\alpha\beta}\psi^v(x,b)+m(\gamma_5)_{\alpha\beta}\psi^s(x,b)],
\end{eqnarray}
where $P$ stands for the momentum of the charmonium meson $\eta_c(2S)$ or $\psi(2S)$ and $m$ is its mass.
The $x$ represents the momentum fraction of the charm quark inside the charmonium, and $b$ is the
conjugate variable of the transverse momentum of the valence quark of the meson.
The $\epsilon^{L(T)}$ denotes its longitudinal (transverse) polarization vector.
The asymptotic models for the twist-2 distribution amplitudes $\psi^{L,T,v}$, and the
twist-3 distribution amplitudes $\psi^{t,V,s}$ will
be derived following the prescription in \cite{epjc60107}.

First, we write down the Schr$\ddot{o}$dinger equal-time wave function
$\Psi_{Sch}(\textbf{r})$ for the harmonic-oscillator potential. The radial wave function of the
corresponding Schr$\ddot{o}$dinger state is given by
\begin{eqnarray}
\Psi_{(2S)}(\textbf{r})\propto(\frac{3}{2}-\alpha^2r^2)e^{-\frac{\alpha^2r^2}{2}},
\end{eqnarray}
where $\alpha^2=\frac{m\omega}{2}$ and $\omega$ is the frequency of oscillations or the quantum of
energy.  We perform the Fourier transformation  to the momentum space to get $\Psi_{2S}({\bf k})$ as
\begin{eqnarray}
\Psi_{(2S)}(\textbf{k})\propto(2k^2-3\alpha^2)e^{-\frac{k^2}{2\alpha^2}},
\end{eqnarray}
with $k^2$ being the square of the three momentum. In terms of the substitution assumption,
\begin{eqnarray}
\textbf{k}_{\perp}\rightarrow \textbf{k}_{\perp},\quad k_z\rightarrow \frac{m_0}{2}(x-\bar{x}),
\quad m_0^2=\frac{m_c^2+\textbf{k}^2_{\perp}}{x\bar{x}},
\end{eqnarray}
with $m_c$ the $c$-quark mass and $\bar{x}=1-x$. We should make the following replacement as regards the
variable $k^2$
\begin{eqnarray}
k^2\rightarrow \frac{\textbf{k}^2_{\perp}+(x-\bar{x})^2m_c^2}{4x\bar{x}}.
\end{eqnarray}
Then the wave function
can be taken as
\begin{eqnarray}
\Psi_{(2S)}(\textbf{k})\rightarrow \Psi_{(2S)}(x,\textbf{k}_{\perp})\propto(\frac{\textbf{k}^2_{\perp}+
m_c^2(x-\bar{x})^2}{2x\bar{x}}-3\alpha^2)e^{-\frac{\textbf{k}^2_{\perp}+
m_c^2(x-\bar{x})^2}{8x\bar{x}\alpha^2}}.
\end{eqnarray}
Applying the Fourier transform to replace the transverse momentum $\textbf{k}_{\perp}$ with its
conjugate variable $b$, the 2S oscillator wave function can be taken as
\begin{eqnarray}\label{eq:wave1}
\Psi_{(2S)}(x,\textbf{b})&\sim& \int \textbf{d}^2\textbf{k}_{\perp}e^{-i\textbf{k}_{\perp}
\cdot \textbf{b}}\Psi_{(2S)}(x,\textbf{k}_{\perp})\nonumber\\
&\propto &
x\bar{x}\mathcal {T}(x)
e^{-x\bar{x}\frac{m_c}{\omega}[\omega^2b^2+(\frac{x-\bar{x}}{2x\bar{x}})^2]},
\end{eqnarray}
with
\begin{eqnarray}
\mathcal {T}(x)=1-4b^2m_c\omega x\bar{x}+\frac{m_c(x-\bar{x})^2}{\omega x\bar{x}}.
\end{eqnarray}
We then propose the 2S states distribution amplitudes
inferred from Eq.~(\ref{eq:wave1}),
\begin{eqnarray}
\Psi_{(2S)}(x,b)
\propto \Phi^{asy}(x)\mathcal {T}(x)
e^{-x\bar{x}\frac{m_c}{\omega}[\omega^2b^2+(\frac{x-\bar{x}}{2x\bar{x}})^2]},
\end{eqnarray}
with the $\Phi^{asy}(x)$ being the asymptotic models, which have been given in \cite{plb612215}.
Therefore, we have the distribution amplitudes for the radially excited charmonium mesons
$\eta_c(2S)$ and $\psi(2S)$
\begin{eqnarray}\label{eq:wave}
\Psi^{L,T,v}(x,b)&=&\frac{f_{2S}}{2\sqrt{2N_c}}N^{L,T,v} x\bar{x}\mathcal {T}(x)
e^{-x\bar{x}\frac{m_c}{\omega}[\omega^2b^2+(\frac{x-\bar{x}}{2x\bar{x}})^2]},\nonumber\\
\Psi^t(x,b)&=&\frac{f_{2S}}{2\sqrt{2N_c}}N^t (x-\bar{x})^2\mathcal {T}(x)
e^{-x\bar{x}\frac{m_c}{\omega}[\omega^2b^2+(\frac{x-\bar{x}}{2x\bar{x}})^2]},\nonumber\\
\Psi^V(x,b)&=&\frac{f_{2S}}{2\sqrt{2N_c}}N^V [1+(x-\bar{x})^2]\mathcal {T}(x)
e^{-x\bar{x}\frac{m_c}{\omega}[\omega^2b^2+(\frac{x-\bar{x}}{2x\bar{x}})^2]},\nonumber\\
\Psi^s(x,b)&=&\frac{f_{2S}}{2\sqrt{2N_c}}N^s \mathcal {T}(x)
e^{-x\bar{x}\frac{m_c}{\omega}[\omega^2b^2+(\frac{x-\bar{x}}{2x\bar{x}})^2]},
\end{eqnarray}
\begin{figure}[!htbh]
\begin{center}
\vspace{-0.5cm}
\centerline{\epsfxsize=7.5 cm \epsffile{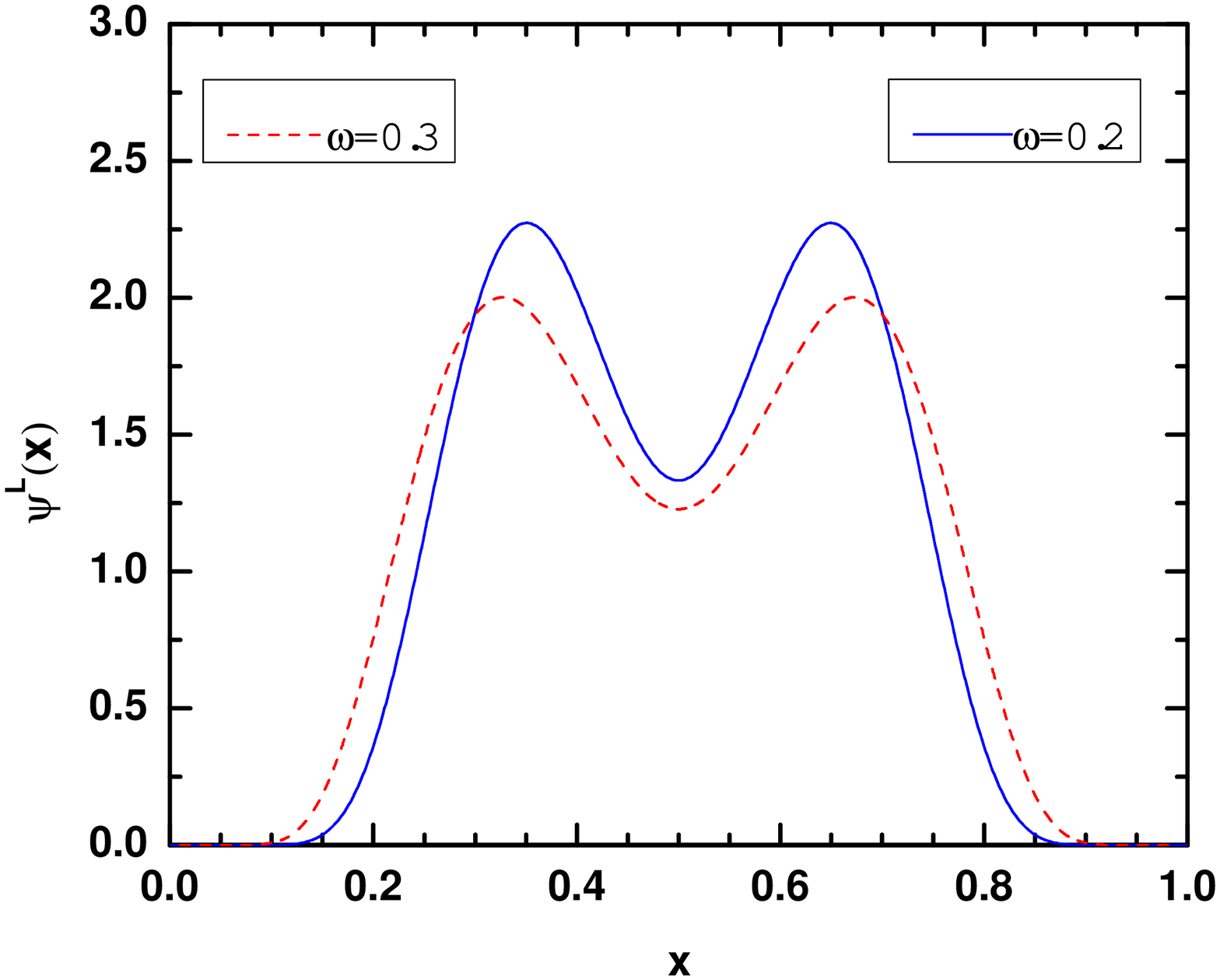}}
\vspace{-0.3cm}
\caption{The shape of the distribution amplitude for $\psi^L(x)$ when $b=0$,
         with the solid (dashed) line for $\omega=0.2(0.3)$ GeV.}
\label{fig:wave}
\vspace{-0.8cm}
\end{center}
\end{figure}
with the normalization conditions:
\begin{eqnarray}
\int_0^1\Psi^{i}(x,0)d x&=&\frac{f_{2S}}{2\sqrt{2N_c}}\;.
\end{eqnarray}
$N_c$ above is the color number,  $N^{i}(i=L,T,t,V,v,s)$ are the normalization constants,
and $f_{2S}$ is the  decay constant of the  2S state. All the distribution amplitudes in Eq.~(\ref{eq:wave})
are symmetric under $x\leftrightarrow \bar{x}$.
Here we do not distinguish the leading twist distribution amplitude  $\Psi^v$ of the  $\eta_c(2S)$
meson from  $\Psi^{L,T}$ of the $\psi(2S)$ meson, and the same decay constant  has been assumed for
the longitudinally and transversely polarized $\psi(2S)$ meson. To make things clearer, the shape of the
distribution amplitude  $\Psi^{L}(x,0)$ is displayed in Fig.~\ref{fig:wave}. The free parameter
$\omega=0.2$  GeV is adopted,
such that the valence charm quark, carrying the invariant mass $x^2P^2\approx m_c^2$, is almost on shell.
It can be seen that the two maximum positions are near $x=0.35$ and $x=0.65$
and a larger value of parameter  $\omega$ gives a wider shape. Note that the dip at $x=0.5$ is a
consequence of the radial Schr$\ddot{o}$dinger wave function of the $n=2,l=0$ state.

\section{form factors and decay amplitudes}

\begin{figure}[!htbh]
\begin{center}
\vspace{-1.0cm}
\centerline{\epsfxsize=12 cm \epsffile{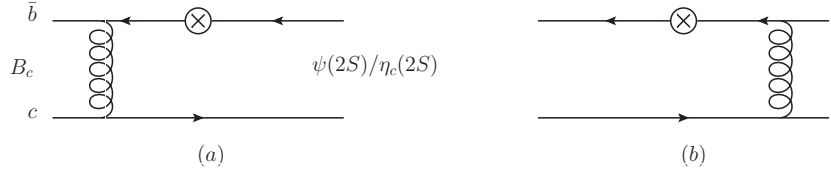}}
\vspace{-10.5cm}\caption{The leading-order Feynman diagrams for the $B_c\rightarrow \left(\psi(2S),
                         \eta_c(2S)\right)$ transitions.}
\label{fig:form}
\vspace{-0.5cm}
 \end{center}
\end{figure}

In the pQCD approach, the $B_c\rightarrow \psi(2S),\eta_c(2S)$ transition form factors in the large-recoil
limit ($q^2=0$), which  are similar to that of $B_c\rightarrow J/\psi,\eta_c$ \cite{cpc37-093102},
can be calculated from above universal hadronic distribution amplitudes.
The lowest-order diagrams  are displayed in Fig.~\ref{fig:form}. The form factors $F_{+,0}(q^2)$,
$V(q^2)$ and $A_{0,1,2}(q^2)$ are defined via the matrix element \cite{prd014009},
\begin{eqnarray}
\label{eq:form1}
\langle \eta_c(2S)(P_2)|\bar{c}\gamma^{\mu}b|B_c(P_1)\rangle&=&\left[(P_1+P_2)^{\mu}
-\frac{M^2-m^2}{q^2}q^{\mu}\right]F_+(q^2) + \frac{M^2-m^2}{q^2}q^{\mu}F_0(q^2),\\
\langle\psi(2S)(P_2)|\bar{c}\gamma^{\mu}b|B_c(P_1)\rangle&=&\frac{2iV(q^2)}{M+m}
\epsilon^{\mu\nu\rho\sigma}\epsilon^*_{\nu}P_{2\rho}P_{1\sigma}\;,\\
\langle\psi(2S)(P_2)|\bar{c}\gamma^{\mu}\gamma_5b|B_c(P_1)\rangle&=&2mA_0(q^2)
\frac{\epsilon^*\cdot q}{q^2}q^{\mu}+ (M+m)A_1(q^2)\left[\epsilon^{*\mu}-
\frac{\epsilon^*\cdot q}{q^2}q^{\mu}\right]\nonumber\\
&-& A_2(q^2)\frac{\epsilon^*\cdot q}{M+m}\left[(P_1+P_2)^{\mu}-\frac{M^2-m^2}{q^2}q^{\mu}\right],
\end{eqnarray}
where $q=P_1-P_2$ is the momentum transfer and $P_1(P_2)$ is the momentum of the initial (final)
state meson. $M$ is the mass of $B_c$ meson, and $\epsilon^*$ is the polarization vector of the
$\psi(2S)$ meson. In the large-recoil limit, say $q^2=0$, we have
\begin{eqnarray}
F_0(0)=F_+(0),\quad A_0(0)=\frac{1+r}{2r}A_1(0)-\frac{1-r}{2r}A_2(0)\;.
\end{eqnarray}
It is straightforward to calculate the form factors $F_{0}(q^2)$,
$V(q^2)$ and $A_{0,1}(q^2)$ at the tree level in the pQCD. They read

\begin{eqnarray}\label{eq:F0ex}
F_0&=&2 \sqrt{\frac{2}{3}} \pi  M^2 f_B C_f\int_0^1dx_2\int_0^{\infty}b_1b_2db_1db_2
\exp\left(-\frac{\omega_B^2 b_1^2}{2}\right)\nonumber\\
&\times&\bigg\{\left[\psi ^v(x_2,b_2) \left(x_2-2 r_b\right)-\psi ^s(x_2,b_2)r \left(2 x_2-r_b\right)\right]
E_{ab}(t_a)h(\alpha_e,\beta_a,b_1,b_2)S_t(x_2)\nonumber\\
&+&\left[\psi ^v(x_2,b_2)(r_c+r^2(1-x_1))-\psi ^s(x_2,b_2)2r(1-x_1+r_c)\right]
E_{ab}(t_b)h(\alpha_e,\beta_b,b_2,b_1)S_t(x_1)\bigg\},
\end{eqnarray}

\begin{eqnarray}\label{eq:Vex}
V&=&2 \sqrt{\frac{2}{3}}(1+r)\pi  M^2 f_B C_f\int_0^1dx_2\int_0^{\infty}b_1b_2db_1db_2
\exp\left(-\frac{\omega_B^2 b_1^2}{2}\right)\nonumber\\
&\times&\bigg\{\left[\psi ^V(x_2,b_2) r\left(1-x_2\right)+\psi ^T(x_2,b_2)(r_b-2)\right]E_{ab}(t_a)h(\alpha_e,\beta_a,b_1,b_2)
S_t(x_2)\nonumber\\
&-&\psi ^V(x_2,b_2)rE_{ab}(t_b)h(\alpha_e,\beta_b,b_2,b_1)S_t(x_1)\bigg\},
\end{eqnarray}
\begin{eqnarray}\label{eq:A0ex}
A_0&=&2 \sqrt{\frac{2}{3}} \pi  M^2 f_B C_f\int_0^1dx_2\int_0^{\infty}b_1b_2db_1db_2
\exp\left(-\frac{\omega_B^2 b_1^2}{2}\right)\nonumber\\
&\times&\bigg\{\left[\psi ^L(x_2,b_2) \left(x_2-2 r_b\right)-\psi ^t(x_2,b_2)r \left(2 x_2-r_b\right)\right]
E_{ab}(t_a)h(\alpha_e,\beta_a,b_1,b_2)S_t(x_2)\nonumber\\
&-&\psi ^L(x_2,b_2)\left[r_c+r^2(1-x_1)\right]E_{ab}(t_b)h(\alpha_e,\beta_b,b_2,b_1)S_t(x_1)\bigg\},
\end{eqnarray}
\begin{eqnarray}\label{eq:A1ex}
A_1&=&2 \sqrt{\frac{2}{3}}\frac{r}{1+r} \pi  M^2 f_B C_f\int_0^1dx_2\int_0^{\infty}b_1b_2db_1db_2
\exp\left(-\frac{\omega_B^2 b_1^2}{2}\right)\nonumber\\
&\times&\bigg\{\left[\psi ^V(x_2,b_2) \left(1+x_2-r^2(1-x_2)-4r_b\right)+\psi ^T(x_2,b_2)\left[r(2-4x_2+r_b)
+\frac{r_b-2}{r}\right]\right]\nonumber\\
&\times&E_{ab}(t_a)h(\alpha_e,\beta_a,b_1,b_2)S_t(x_2)
-\psi ^V(x_2,b_2)\left[1-2x_1+2r_c+r^2\right]E_{ab}(t_b)h(\alpha_e,\beta_b,b_2,b_1)S_t(x_1)\bigg\},
\end{eqnarray}
with $r=\frac{m}{M}$ and $r_{b,c}=\frac{m_{b,c}}{M}$. The functions $E_{ab}$, the scales $t_{a,b}$
and the hard functions $h$ are given in Appendix B of Ref. \cite{14075550}.

\begin{figure}[] %%!htbh
\begin{center}
\vspace{-1cm} \centerline{\epsfxsize=9.5 cm \epsffile{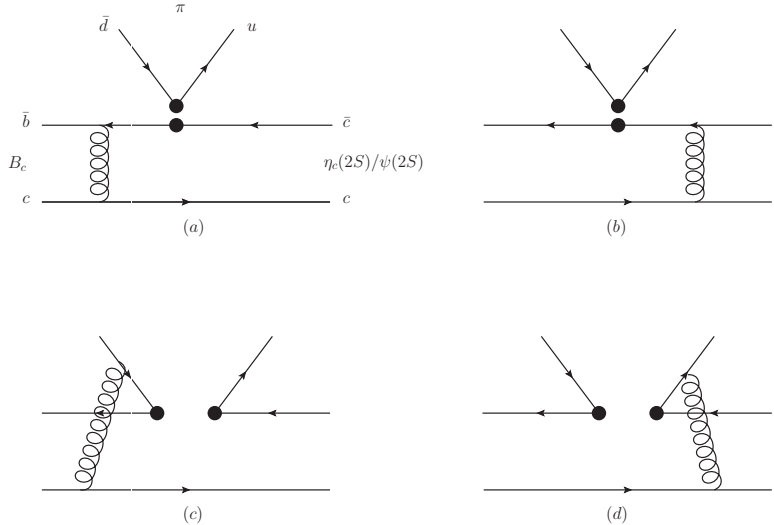}}
\vspace{-5.0cm} \caption{Feynman diagrams for  $B_c\rightarrow \psi(2S)\pi,\eta_c(2S)\pi$ decays.}
\label{fig:lo}
\vspace{-1.0cm}
\end{center}
\end{figure}

The quark diagrams contributing to the $B_c\rightarrow \psi(2S)\pi,\eta_c(2S)\pi$ decays are
displayed in Fig.~\ref{fig:lo},
where (a) and (b) are for the factorizable topology, and (c) and (d) are for the nonfactorizable topology.
The effective Hamiltonian relevant to the considered decays  is written as \cite{rmp681125}
\begin{eqnarray}
\mathcal{H}_{eff}=\frac{G_F}{\sqrt{2}}V^*_{cb}V_{ud}[C_1(\mu)O_1(\mu)+C_2(\mu)O_2(\mu)]+\text{h.c.},
\end{eqnarray}
with $V^*_{cb}$ and $V_{ud}$  the Cabibbo-Kobayashi-Maskawa (CKM) matrix elements, $C_{1,2}(\mu)$
the Wilson coefficients, and $O_{1,2}(\mu)$ the effective four quark operators
\begin{eqnarray}
O_{1}(\mu)&=&\bar{b}_{\alpha}\gamma^{\mu}(1-\gamma_5)c_{\beta} \otimes \bar{u_{\beta}}\gamma_{\mu}
(1-\gamma_5)d_{\alpha},\nonumber\\
O_{2}(\mu)&=&\bar{b}_{\alpha}\gamma^{\mu}(1-\gamma_5)c_{\alpha}\otimes \bar{u}_{\beta}\gamma_{\mu}
(1-\gamma_5)d_{\beta},
\end{eqnarray}
where $\alpha$ and $\beta$ are the color indices. Since the four quarks in the operators are different
from each other, there is no penguin contribution. Therefore there will be  no $CP$ violation in the decays of
$B_c\rightarrow \psi(2S)\pi,\eta_c(2S)\pi$ within the standard model. After a straightforward calculation using the
pQCD formalism of Eq.~(\ref{eq:factorization}), we have the decay amplitudes
\begin{eqnarray}\label{eq:amplitude1}
\mathcal {A}(B_c\rightarrow (\psi(2S), \eta_c(2S))\pi)&=&V_{cb}^*V_{ud}[(C_2+\frac{1}{3}C_1)
\mathcal {F}_{e}+C_1\mathcal {M}_{e}].
\end{eqnarray}
The detailed expressions of $\mathcal {F}_{e}$ and $\mathcal {M}_{e}$  are the same as the
$B_c\rightarrow (J/\psi, \eta_c)\pi$ decay modes in Appendix A of Ref.~\cite{14075550},
except for the replacements $J/\psi\rightarrow\psi(2S)$ and $\eta_c\rightarrow\eta_c(2S)$.

\section{Numerical results and discussions}\label{sec:num}

In the numerical calculations we need the following input parameters (in units of GeV)~\cite{pdg}:
\begin{eqnarray}
m_{c}=1.275,\;\; m_b=4.18,\;\; M_{B_c}=6.277,\;\; m_{\psi(2S)}=3.686,\;\; m_{\eta_c(2S)}=3.639.
\end{eqnarray}
For the relevant CKM matrix elements we use $V_{cb}=(40.9 \pm 1.1)\times 10^{-3}$ and
$V_{ud}=0.97425\pm 0.00022$~\cite{pdg}.

The decay constant $f_{\psi(2S)}$ can be derived from the process  $\psi(2S)\rightarrow e^+e^-$
by the relationship
\begin{eqnarray}\label{eq:fpsi}
f_{\psi(2S)}=\sqrt{\frac{3m_{\psi(2S)}\Gamma_{\psi(2S)\rightarrow e^+e^-}}{4\pi\alpha^2Q_c^2}},
\end{eqnarray}
using the  data given in \cite{pdg}
\begin{eqnarray}
\Gamma_{\psi(2S)\rightarrow e^+e^-}=(2.36\pm0.04)  \quad \text{keV}.
\end{eqnarray}
Then we   have $f_{\psi(2S)}=296^{+3}_{-2}$ MeV. The decay constant $f_{\eta_c(2S)}$ can be
determined by the double photon decay of $\eta_c(2S)$ as
\begin{eqnarray}\label{eq:feta}
f_{\eta_c(2S)}=\sqrt{\frac{81\pi m_{\eta_c(2S)}\Gamma_{\eta_c(2S)\rightarrow \gamma\gamma}}
{4(4\pi\alpha)^2}}.
\end{eqnarray}
 Using the measured results of the branching fractions  $\eta_c(2S)\rightarrow \gamma\gamma$ and the
full width of $\eta_c(2S)$ \cite{pdg},
\begin{eqnarray}\label{eq:eee}
\mathcal {B}(\eta_c(2S)\rightarrow \gamma\gamma)=(1.9\pm1.3)\times 10^{-4},\;\;
\Gamma_{\eta_c(2S)}=11.3^{+3.2}_{-2.9} \text{MeV},
\end{eqnarray}
we can get the decay constant $f_{\eta_c(2S)}=243^{+79}_{-111}$ MeV.
As for the decay constant for $B_c$, we adopt $f_{B_c}=489$ MeV~\cite{plb651-171}.

\begin{table}
\caption{The form factors for $F_0^{B_c\rightarrow \eta_c(2S)}$, $A^{B_c\rightarrow \psi(2S)}_{0,1,2}$ and
$V^{B_c\rightarrow \psi(2S)}$ at $q^2=0$ evaluated by pQCD and by other methods in the literature. We also show theoretical uncertainties induced by the shape parameters,  $m_c$, $f_{\psi(2s)}$ or $f_{\eta_c(2s)}$ and
the hard scale $t$, respectively.}
\label{tab:formfactor}
\begin{tabular}[t]{l|c|c|c|c|c}
\hline\hline
%\toprule[2pt]
&This work&Ref.\cite{prd017501}&Ref.\cite{11022190}\footnotemark[1]&Ref.\cite{prd094020}&Ref.\cite{prd054003}\\ \hline
  $F_0$ & $0.70^{+0.09+0.12+0.23+0.02}_{-0.05-0.10-0.32-0.01}$&-- &0.325&0.27&--\\
   $A_0$   &$0.56^{+0.09+0.07+0.00+0.01}_{-0.05-0.04-0.00-0.01}$   &0.45&0.42&0.23&0.20\\
   $A_1$    &$0.56^{+0.13+0.06+0.00+0.01}_{-0.04-0.03-0.00-0.01}$   &0.335 &0.35&0.18&0.38 \\
   $A_2$    &$0.62^{+0.27+0.04+0.01+0.02}_{-0.05-0.01-0.01-0.00}$&0.102 &0.15&0.14&0.90\\
   $V_0$     &$0.95^{+0.18+0.15+0.01+0.03}_{-0.08-0.10-0.01-0.01}$&0.525&0.73&0.24&0.90\\
%\bottomrule[2pt]
\hline\hline
\end{tabular}
\footnotetext[1]{ Comparing the definitions of the transition form factor of
Ref. \cite{11022190} with ours, we have the following relations at the maximal recoil point:
\begin{eqnarray}
F_0&=&f^+,\quad V=(M+m)g,\quad A_1=\frac{f}{M+m},\quad A_2=-(M+m)a_+,\nonumber\\ A_0&=&\frac{f+(M^2-m^2)a_++q^2a_-}{2m},\nonumber
\end{eqnarray}
where the values of $f^+,g,f,a_+,a_-$ are given in \cite{11022190}.}
\end{table}

Our numerical results for the form factors $F_0^{B_c\rightarrow \eta_c(2S)}$,
$A^{B_c\rightarrow \psi(2S)}_{0,1,2}$ and $V^{B_c\rightarrow \psi(2S)}$
are listed in Table \ref{tab:formfactor}.
We find that the form factors are close by different approaches within errors, except the
results in Ref. \cite{prd094020} which are typically smaller. Some dominant uncertainties are
considered in our numerical values: the first error comes from the shape parameters
$\omega_B=0.6\pm0.1$ ($\omega=0.2\pm0.1$) GeV for the $B_c$($\psi(2S)/\eta_c(2S)$) meson, the second
one is induced by $m_c=1.275\pm0.025$ GeV, the third error comes from the decay constants of
the $\psi(2S)$ or $\eta_c(2S)$ meson, and the last one is caused by the variation of  the hard scale
from $0.75t$ to $1.25t$ in  Eq.~(\ref{eq:factorization}), which characterizes the size of
next-to-leading-order contribution. It is found that the main errors
come from the uncertainties of the shape parameters and the charm-quark mass.
Therefore, the decay  of $B_c\rightarrow\psi(2S)(\eta_c(2S))$  provides a good platform to understand
the wave function of the radially excited charmonium states and the constituent quark
model. The uncertainty from the decay constant of $\eta_c(2S)$
meson is large due to the low accuracy measurement of the  branching fraction
in Eq.~(\ref{eq:eee}); the relevant uncertainty of $F_0$  is large, too. We expect that it could be
measured precisely  at LHCb and  Super-B factories in the near future. We also noticed that the
error from the uncertainty of the hard scale $t$ is small, which means the next-to-leading-order
contributions can be safely neglected. The errors from the uncertainty of the CKM matrix elements are
very small and they have been neglected.

\begin{table}
\caption{Branching ratios ($10^{-4}$) of the $B_c\rightarrow \eta_c(2S)\pi, \psi(2S)\pi$ decays.
The errors induced by the same sources as they in Table~\ref{tab:formfactor}.}
\label{tab:br}
\begin{tabular}[t]{l c c c c c c c c c}
\hline\hline
%\toprule[2pt]
Modes     & This work & \cite{prd017501}\footnotemark[1]&\cite{11022190}&\cite{prd094020}&\cite{prd4133}
&\cite{prd3399}&\cite{prd034012}
&\cite{14113428}&\cite{prd034008}\footnotemark[2]
 \\ \hline
$B_c\rightarrow \eta_c(2S)\pi$ \; &\; $10.3^{+3.4+4.0+7.8+1.2}_{-1.8-2.8-7.2-0.4}$\;\;
&--&2.4&1.7 &2.2&2.4&0.66&2.87&--\\
$B_c\rightarrow \psi(2S)\pi$   & $6.7^{+2.8+1.8+0.1+0.7}_{-1.1-1.2-0.1-0.3}$
&2.97&3.7& 1.1 &0.63&2.2&2.0&2.66&7.6(5.8)\\
%\bottomrule[2pt]
\hline\hline
\end{tabular}
\footnotetext[1]{We quote the result with the modified wave functions  for $\psi(2S)$.}
\footnotetext[2]{The nonbracketed (bracketed) results are evaluated at the NLO (LO) level.}
\end{table}

The branching fractions for the $B_c\rightarrow \eta_c(2S)\pi, \psi(2S)\pi$ decays in the $B_c$ meson
rest frame can be written as
\begin{eqnarray}
\mathcal {B}(B_c\rightarrow(\psi(2S), \eta_c(2S))\pi)&=&\frac{G_F^2\tau_{B_c}}{32\pi M_B}(1-r^2)|
\mathcal {A}|^2,
\end{eqnarray}
where the decay amplitudes $\mathcal {A}$  have been given explicitly in Eq.~(\ref{eq:amplitude1}).
In Table~\ref{tab:br}, we show the results of the branching fractions for the two-body nonleptonic
$B_c\rightarrow \eta_c(2S)\pi, \psi(2S)\pi$ decays, where the sources of the errors in the
numerical estimates have the same origin as in the discussion of the form factors in
Table \ref{tab:formfactor}.
It is easy to see that the most important theoretical uncertainties are caused by the nonperturbative
shape parameters, the charm-quark mass, and the decay constant $f_{\eta_c(2S)}$, which can be improved
by future experiments. It is found that the branching fractions of $B_c$ decays to the 2S  state are smaller
than those of 1S state in our previous study \cite{14075550} in the perturbative QCD approach.
This phenomenon can be understood from the wave functions of the two states. The presence of the node
in the 2S wave function, which can be seen in Fig.~\ref{fig:wave},  causes the overlap
between the initial and final state wave functions to becomes smaller.  Besides,  the tighter
phase space and the smaller decay constants of 2S state  also  suppress their branching ratios.

We also make a comparison of our results with the previous studies. One can see that our results are
comparable to those of  \cite{prd034008} within the error bars, but larger than the results from other
modes. This is because  they have used the  smaller form factors at maximum recoil. Regardless of
this effect, our results are consistent with theirs. For example, as shown in Tables
\ref{tab:formfactor} and \ref{tab:br}, our values of $A_0$ and $F_0$  are about $2.5$ times
the results of Ref. \cite{prd094020}, and result in our branching ratios are $6$  times larger than
theirs. For a more direct comparison with the available experimental data, we compare the present
results in Table \ref{tab:br} with those for the decays of $B_c$ to S-wave charmonium states $J/\psi$
and $\eta_c$ (also based on the harmonic-oscillator wave functions), whose results can be
found in Ref. \cite{14075550}, and obtain the  ratios $\mathcal {B}(B_c\rightarrow(\psi(2S) \pi))/
\mathcal {B}(B_c\rightarrow(J/\psi \pi))=0.29^{+0.17}_{-0.11}$ and
$\mathcal {B}(B_c\rightarrow(\eta_c(2S) \pi))/\mathcal {B}(B_c\rightarrow(\eta_c \pi))
=0.35^{+0.36}_{-0.29}$.
The former is consistent with the data $0.25\pm0.068\pm0.014$ \cite{prd071103}, and also comparable
with the recent prediction of the Bethe-Salpeter relativistic quark model \cite{14113428}, $0.24$.
This fact may indicate that the  harmonic-oscillator wave functions for radially excited states are
reasonable and applicable.  Although the $B_c\rightarrow \eta_c(2S)\pi$  decay  has not yet been
measured so far,  the predicted large branching ratio ($10^{-3}$) makes it possible to   measure it soon   at the LHCb
 experiment or future facility.

\begin{table}
\caption{The values of decay amplitude from twist-2 and twist-3 charmonium wave functions for
$B_c\rightarrow \eta_c(2S)\pi, \psi(2S)\pi$ decays. The results are given in units of $\text{GeV}^3$.}
\label{tab:amp}
\begin{tabular}[t]{l c c c}
\hline\hline
Modes     &twist-2&twist-3&total \\ \hline
$\mathcal {A}(B_c\rightarrow \psi(2S)\pi)$\;\; & -1.7-0.07i\;\;&-0.4-0.06i\;\;&-2.1-0.13i\\
$\mathcal {A}(B_c\rightarrow \eta_c(2S)\pi)$\;\; & -1.5-2.3i\;\;&3.9+1.4i\;\;&2.4+0.9i\\
\hline\hline
\end{tabular}
\end{table}

We now investigate the relative importance of the twist-2
and twist-3 contributions in Eq.~(\ref{eq:wavve}) to the decay amplitude, whose   results are
 displayed separately in Table \ref{tab:amp}, where the label ``twist-2 (twist-3)"
corresponds to the contribution of the twist-2 (twist-3) distribution amplitude
only, while the label ``total" corresponds to both of the contributions. It is found that the
contribution of  twist-3 distribution amplitude
is not power-suppressed for $B_c\rightarrow \eta_c(2S)\pi$ decay, whose   contribution  is $1.5$ times
larger than the twist-2 contribution. The reason  is that the term $\psi ^s(x_2,b_2)2r$ in
Eq.~(\ref{eq:F0ex}) from Fig.~\ref{fig:lo} b gives the dominant contribution to the decay amplitude, since
the asymptotic model of the twist-3 distribution amplitude in Eq.~(\ref{eq:wave}) for the $\eta_c(2S)$ meson has no
factor like $x(1-x)$ to suppress its integral value in the end-point region, which
leads to large enhancement compared with twist-2 contribution. However, because twist-3 terms of the
$\psi(2S)$ meson distribution amplitude do not contribute to the $B_c\rightarrow \psi(2S)\pi$ decay
amplitude from Fig.\ref{fig:lo} b, the contribution from other diagrams with  twist-3 distribution
amplitude is only one-fifth smaller than that of the twist-2 contribution in this process.
It is also found that there is very strong interference between contributions of the twist-2 and twist-3
wave functions  for both  $B_c\rightarrow\psi(2S)\pi$ and $B_c\rightarrow \eta_c(2S)\pi$ decays.
The numerical results show that the contributions from twist-3 wave function
have an opposite sign between the two  channels.
This results in  constructive interference for the former, but  destructive interference for the
latter. The reason is that the amplitudes are different between the two decays
at twist-3 level, which can be seen in Eqs. (A1) and (A4) of \cite{14075550}. A similar situation also
exists in $B_c\rightarrow D\pi,D^*\pi$ \cite{prd074008} decays.

\section{ conclusion}
We calculated the form factors of the weak $B_c$ decays to radially excited charmonia and the
branching ratios of $B_c\rightarrow \psi(2S)\pi,\eta_c(2S)\pi$ decays in the pQCD approach.
The new  charmonium  distribution amplitudes based on the radial Schr$\ddot{o}$dinger wave function of the
$n=2,l=0$ state for the  harmonic-oscillator potential are employed. We discussed
theoretical uncertainties arising from the nonperturbative shape parameters, the charm-quark mass,
the decay constants,  and the scale dependence. It is found that the main uncertainties of the processes
concerned come from the shape parameters and the charm-quark mass. The theoretically
evaluated ratio $\mathcal {B}(B_c\rightarrow(\psi(2S) \pi))
/\mathcal {B}(B_c\rightarrow(J/\psi \pi))=0.29^{+0.17}_{-0.11}$ is consistent with the data, which
indicates the  harmonic-oscillator wave functions work well, not only for the
ground state charmonium, but also for the radially excited charmonia.  It is also found that the
twist-3 charmonium distribution amplitude  gives a large contribution, especially
for $B_c\rightarrow\eta_c(2S)\pi$ decay, whose  branching fraction is of the
order of  $10^{-3}$, which could be tested at the ongoing large hadron collider.

\begin{acknowledgments}
This work is supported in part by the National Natural Science Foundation of China under
Grants No. 11235005, No. 11347168, 11375208, and No. 11405043, by the Natural Science Foundation of
Hebei Province of China under Grant No. A2014209308, and by the China Postdoctoral Science Foundation.

\end{acknowledgments}

%%\begin{appendix}
%%\end{appendix}
%%%%%%%%%%%%%%%%%%%%%%%%%%%%%%%%%%%%%%%%%%%%%%%%%%%%%%%%%%%%%%%%%%

\end{document}